\documentclass[prd,aps,showpacs,preprintnumbers,amssymb]{revtex4}
\usepackage{axodraw}
\usepackage{color}
\usepackage{epsf}

\def\e3p{$\eta \rightarrow 3 \pi$}

\begin{document}
\title{%
\hfill{\normalsize\vbox{%
\hbox{}
 }}\\
{A hierarchy of the quark masses in a top condensate model with multiple Higgses}}

\author{Amir H. Fariborz
$^{\it \bf a}$~\footnote[1]{Email:
 fariboa@sunyit.edu}}
\author{Renata Jora
$^{\it \bf a}$~\footnote[2]{Email:
 rjora@theory.nipne.ro}}
\author{Salah Nasri$^{\it \bf b}$~\footnote[3]{Email:
 nasri.salah@gmail.com}}
\author{Joseph Schechter
 $^{\it \bf c}$~\footnote[4]{Email:
 schechte@phy.syr.edu}}

\affiliation{$^{\bf \it a}$ Department of Engineering, Science and Mathematics, State University of New York Institute of Technology, Utica, NY 13504-3050, USA}
\affiliation{$^{\bf \it b}$ National Institute of Physics and Nuclear Engineering PO Box MG-6, Bucharest-Magurele, Romania}
\affiliation{$^{\bf \it c}$ Department of Physics, College of Science, United Arab Emirates University, Al-Ain, UAE}

\affiliation{$^ {\bf \it d}$ Department of Physics,
 Syracuse University, Syracuse, NY 13244-1130, USA}

\date{\today}

\begin{abstract}
We discuss the quark masses in a top condensate model where not only two quark but also four quark composite states may exist. We show that the presence of the top color
group $SU(3)_1\times SU(3)_2$ with the correct quark representations can justify even in the absence of the additional technicolor interactions a hierarchy
of the quark masses where the light quarks masses have the same size, the charm and bottom masses are higher and similar and the top is the heaviest.
\end{abstract}
\maketitle

\section{Introduction}
One of the interesting candidates for a theory with dynamical symmetry breaking is the top condensate model with its various extensions \cite{Lindner}-\cite{Hill3}. In \cite{Jora} we presented  a top condensate model  from the perspective considered in low energy QCD that not only two quark states but also four quark states might exist. Then the strong dynamics suggested that  the Higgs like particle discovered at the LHC might correspond to the scalar from an electroweak triplet of four quark states composed of the top and bottom quarks. To make the model consistent one assumes an extra strong top color interaction and also other additional particles and symmetries.
Here we would like to discuss the possibility of giving masses in this type of model to  a part of the fermions in the standard model in a minimal way without involving any elementary scalars or additional interactions. For that we will need to modify the set-up depicted in \cite{Jora} to include more Higgs states of two or four quark nature although  we  assume that the main part of the findings there are still valid. Our purpose here is however to show that the model can generate the correct hierarchy of masses for the quarks.
We will delegate the problem of the Higgs states and that of the lepton masses to another work.

The paper is organized as follows.  We show in section II that the higher dimensional operators with standard model interactions fail to generate the correct size for the masses of the light quarks. In section III we briefly review some aspects of low energy QCD where  four quark states might help to explain the spectroscopy of light scalar and pseudoscalar mesons.  Section IV contains an outline of the top condensate model proposed in \cite{Jora} which contains both two top and four top composite states. In section V we briefly discuss the basis of the top color theories and in section VI we consider our version of the model. In section VII we present a summary and discuss in detail how our model generates small masses for the light quarks (up, down and strange), higher masses for the  charm and bottom quarks and an even higher mass for the top quark.


\section{Four fermion interactions in the standard model}

The standard model is the most obvious example of a theory  which below the scale of the Z and W bosons contains  four fermion interactions. It is known that quarks form condensates when the color interaction becomes strong enough so the neutral-neutral currents interactions have the correct structure  to contribute to the masses through this mechanism.

For simplicity we will start by listing the four fermion diagonal interactions terms in the standard model. The corresponding Lagrangian is:
\begin{eqnarray}
&&{\cal L}_{eff}=
\frac{-g^2}{2cos^2 \theta_W M_Z^2}J^0_{\mu}J^{0\mu}
\nonumber\\
&&J_{\mu}^0=\sum[g_L^f \bar{f}_L\gamma^{\mu}f_L+g_R^f \bar{f}_R\gamma^{\mu}f_R]
\label{f5645}
\end{eqnarray}

We will select from these only those terms which by fermion condensation may lead to mass terms, explicitly:
\begin{eqnarray}
-\frac{g^2}{cos^2\theta_W M_Z^2}[\sum g_L^f \bar{f}_L\gamma^{\mu}f_L][\sum g_R^f \bar{f}_R\gamma^{\mu} f_R]
\label{somet5546}
\end{eqnarray}

We will list below the four situations of interest in the quark sector. We denote a generic up quark by $q_u$ and down quark by $q_d$):
\begin{eqnarray}
&&-\frac{g^2}{cos^2\theta_W M_Z^2}(1/2-2/3sin^2\theta_W)(1/3 sin^2\theta_W) \bar{q}_{uL}\gamma^{\mu}q_{uL}\bar{q}_{dR}\gamma^{\mu}q_{dR}
\nonumber\\
&&-\frac{g^2}{cos^2\theta_W M_Z^2}(-1/2+1/3sin^2\theta_W)(-2/3sin^2\theta_W)\bar{q}_{dL}\gamma^{\mu}q_{dL}\bar{q}_{uR}\gamma^{\mu}q_{uR}
\nonumber\\
&&-\frac{g^2}{cos^2\theta_W M_Z^2}(1/2-2/3sin^2\theta_W)(-2/3sin^2\theta_W)\bar{q}_{uL}\gamma^{\mu}q_{uL}\bar{q}_{uR}\gamma^{\mu}q_{uR}
\nonumber\\
&&-\frac{g^2}{cos^2\theta_W M_Z^2}(-1/2+1/3sin^2\theta_W)(1/3sin^2\theta_W)\bar{q}_{dL}\gamma^{\mu}q_{dL}\bar{q}_{dR}\gamma^{\mu}q_{dR}.
\label{list66577}
\end{eqnarray}

In the same order we list the leading contribution after a Fierz rearrangement of the terms in Eq. (\ref{list66577}):
\begin{eqnarray}
&&\frac{g^2}{cos^2\theta_W M_Z^2}(1/2-2/3sin^2\theta_W)(1/3)sin^2\theta_W)\bar{q}_{u}q_d\bar{q}_d q_u+...
\nonumber\\
&&\frac{g^2}{cos^2\theta_W M_Z^2}(-1/2+1/3sin^2\theta_W)(-2/3sin^2\theta_W)\bar{q}_d q_u\bar{q}_u q_d+...
\nonumber\\
&&\frac{g^2}{cos^2\theta_W M_Z^2}(1/2-2/3sin^2\theta_W)(-2/3sin^2\theta_W)\bar{q}_u q_u\bar{q}_u q_u+...
\nonumber\\
&&\frac{g^2}{cos^2\theta_W M_Z^2}(-1/2+1/3sin^2\theta_W)(1/3sin^2\theta_W)\bar{q}_d q_d\bar{q}_d q_d+....
\label{res554664}
\end{eqnarray}

First a short look at the Eq. (\ref{res554664}) shows that  a light quark condensate leads to a mass term for the quarks proportional to the chiral condensate such that:
\begin{eqnarray}
&&m_{q_u}=-\frac{g^2}{cos^2\theta_W M_Z^2}(1/2-2/3sin^2\theta_W)(-2/3sin^2\theta_W)\frac{1}{\Lambda^2}\langle \bar{q}_u q_u\rangle
\nonumber\\
&&m_{q_d}=-\frac{g^2}{cos^2\theta_W M_Z^2}(-1/2+1/3sin^2\theta_W)(2/3sin^2\theta_W)\frac{1}{\Lambda^2}\langle\bar{q_d}q_d \rangle.
\label{res554554}
\end{eqnarray}

For a light quark vacuum condensate  of $\langle|\bar{q}q|\rangle\cong-0.016$ GeV these masses become:
\begin{eqnarray}
&&m_u=-4.31\times 10^{-7}\,\,GeV
\nonumber\\
&&m_d=-2.63\times 10^{-7}\,\,GeV.
\label{mass56477}
\end{eqnarray}

This shows that although the mechanism is right one cannot obtain only with the standard model interactions the correct size of the quark masses.

\section{Review of the QCD analogue}

It is known that the light scalar mesons can be fit into nonets \cite{Jora1} with regard to the global chiral group $SU(3)_L\times SU(3)_R$.  The low energy spectroscopy of QCD suggests that these nonets might be an admixture of two quark states with the schematic realization,
\begin{eqnarray}
M_a^b=(q_{bA})^{\dagger}\gamma^4\frac{1+\gamma_5}{2}q_{aA},
\label{tw5534}
\end{eqnarray}

and four quark ones for which there are three possibilities. The first one is that the four quark states are molecules made out of two quark-antiquark fields:
\begin{eqnarray}
M^{(2)b}_a=\epsilon_{acd}\epsilon^{bef}(M^{\dagger})_e^c(M^{\dagger})_f^d
\label{four6657}
\end{eqnarray}

Another possibility is that the four quark structures may be bound states of a diquark and anti-diquark. Here there are two choices.
In the first case the diquark is in ${\bar 3}$ of flavor, ${\bar 3}$ of color and has spin zero:
\begin{eqnarray}
&&L^{gE}=\epsilon^{gab}\epsilon^{EAB}q^T_{aA}C^{-1}\frac{1+\gamma_5}{2}q_{bB}
\nonumber\\
&&R^{gE}=\epsilon^{gab}\epsilon^{EAB}q^T_{aA}C^{-1}\frac{1-\gamma_5}{2}q_{bB}
\label{repr45554}
\end{eqnarray}

The matrix M has the structure:

\begin{eqnarray}
M_g^{(3)f}=(L^{gA})^{\dagger}R^{fA}
\label{res4443}
\end{eqnarray}

In the second case the diquark is in ${\bar 3}$ of flavor, 6 of color and has spin 1:
\begin{eqnarray}
&&L^g_{\mu\nu,AB}=L^g_{\mu\nu,BA}=\epsilon^{gab}Q^T_{aA}C^{-1}\sigma_{\mu\nu}\frac{1+\gamma_5}{2}Q_{bB}
\nonumber\\
&&R^g_{\mu\nu, AB}=R^g_{\mu\nu,BA}=\epsilon^{gab}Q^T_{aA}C^{-1}\sigma_{\mu\nu}\frac{1-\gamma_5}{2}Q_{bB}.
\label{res5554546}
\end{eqnarray}

Here $\sigma_{\mu\nu}=\frac{1}{2i}[\gamma_{\mu},\gamma_{\nu}]$. The matrix M has the form:
\begin{eqnarray}
M_g^{(4)f}=(L^g_{\mu\nu,AB})^{\dagger}R^f_{\mu\nu,AB}
\label{res5553443}
\end{eqnarray}

It can be shown using Fierz transformations that the three four quark structures are actually linearly dependent.
We shall consider here for simplicity only the structures in Eq. (\ref{four6657}).


\section{An outline of a top condensate model with a Higgs doublet and a Higgs triplet}

In \cite{Jora} we considered a top condensate model where not only two quark states but also four quark structures contribute to the dynamical breaking of the $U(1)\times SU(2)_L$ symmetry.
These fields correspond to a two Higgs doublet:
\begin{eqnarray}
\Phi_1=
\left[
\begin{array}{c}
t^{\dagger}_R b_L\\
t^{\dagger}_R t_L
\end{array}
\right].
\label{doubl56665}
\end{eqnarray}
and to two Higgs triplets, one with $Y=2$,

\begin{eqnarray}
&&\chi^{++}=n b_L^{\dagger}t_R b_L^{\dagger}t_R
\nonumber\\
&&\chi^+=n b_L^{\dagger}t_Rt_L^{\dagger}t_R
\nonumber\\
&&\chi^0=n t_L^{\dagger}t_Rt_L^{\dagger}t_R,
\label{trpl88796}
\end{eqnarray}
and the other one  with hypercharge $Y=0$,

\begin{eqnarray}
&&\xi^0=n (t_R^{\dagger}t_L)(t_R^{\dagger}t_L)^{\dagger}
\nonumber\\
&&\xi^+=n (t_R^{\dagger}t_L)(t_R^{\dagger}b_L)^{\dagger}
\nonumber\\
&&\xi^-=n (t_R^{\dagger}b_L)(t_R^{\dagger}t_L)^{\dagger}.
\label{tr77564}
\end{eqnarray}

The full Higgs triplet then takes the standard form \cite{Georgi}:
\begin{eqnarray}
\chi=
\left[
\begin{array}{ccc}
\chi^0&\xi^+&\chi^{++}\\
\chi^{-}&\xi^0&\chi^+\\
\chi^{--}&\xi^{-}&\chi^0
\end{array}
\right]
\label{full6575}
\end{eqnarray}

The fields  in the model develop the vev's:
\begin{eqnarray}
&&\langle\Phi_0\rangle=\frac{a}{\sqrt{2}}
\nonumber\\
&&\langle\chi^0\rangle=b
\nonumber\\
&&\langle\xi^0\rangle=b
\label{veve443535}
\end{eqnarray}
and the following relations consistent with the electroweak symmetry breaking hold:

\begin{eqnarray}
&&v^2=a^2+8b^2
\nonumber\\
&&c_H=\frac{a}{a^2+8b^2}
\nonumber\\
&&s_H=[\frac{8b^2}{a^2+8b^2}]^{1/2}.
\label{rel7756454}
\end{eqnarray}

After spontaneous symmetry breakdown the model contains the charged states (according to the classification under the SU(2) custodial symmetry \cite{Georgi}):

\begin{eqnarray}
&&H_5^{++}=\chi^{++}
\nonumber\\
&&H_5^+=\frac{1}{\sqrt{2}}(\chi^+-\xi^+)
\nonumber\\
&&H_3^+=\frac{a(\chi^{+}+\xi^+)-4b\Phi^+}{\sqrt{2}(a^2+8b^2)^{1/2}}
\label{st44335}
\end{eqnarray}
and the neutral states:
\begin{eqnarray}
&&H_5^0=\frac{1}{\sqrt{6}}(2\xi^0-\chi^0-\chi^{0*})
\nonumber\\
&&H_3^0=\frac{a(\chi^0-\chi^{0*})-2\sqrt{2}b(\Phi^0-\Phi^{0*})}{\sqrt{2}(a^2+8b^2)^{1/2}}
\nonumber\\
&&H_1=\frac{1}{\sqrt{2}}(\Phi_0+\Phi_0*)
\nonumber\\
&&H_1'=\frac{1}{\sqrt{3}}(\chi^0+\chi^{0*}+\xi^0).
\label{st4432}
\end{eqnarray}

For simplicity we adopted in \cite{Jora} the point of view that the Higgs doublet and the Higgs triplet do not mix with each other. In this situation the strong dynamics suggests that the Higgs boson found at the LHC should be identified with the neutral scalar corresponding to the Higgs triplet.

\section{Brief review of the simplest topcolor model}

The top color models assume the existence of a larger topcolor group \cite{Hill} $SU(3)_1\times SU(3)_2$. This group is broken at scale M by a scalar field which may or may not be composite down to the color group $SU(3)_C$.
The relation between gauge fields $A^C_{1\mu}$ and $A^C_{2\mu}$ and the coloron ($B_{\mu}^A$) and color ($A_{\mu}^A$) gauge fields is given by:
\begin{eqnarray}
&&A^C_{1\mu}=\cos\theta A_{\mu}^C-\sin\theta B_{\mu}^C
\nonumber\\
&&A^C_{2\mu}=\sin\theta A_{\mu}^C+\cos\theta B_{\mu}^C
\label{res33325}
\end{eqnarray}
where,
\begin{eqnarray}
h_1\cos\theta=g_3\,\,\,\,h_2\sin\theta=g_3.
\label{expr55667}
\end{eqnarray}

Here $h_1$ and $h_2$ are the gauge coupling constants corresponding to the groups $SU(3)_1$ and $SU(3)_2$ respectively whereas $g_3$ is the QCD gauge coupling constant. The values $h_1$ and $h_2$ are chosen such that $h_2\gg h_1$ and $\cot\theta=\frac{h_2}{h_1}\gg1 $.
Upon integrating out the heavy colorons with mass $M_B=\sqrt{h_1^2+h_2^2}M$ one obtains the four fermion interactions:
\begin{eqnarray}
{\cal L}_{eff}=-\frac{1}{M_B^2}h^A_{\mu}h^{A\mu}
\label{int6657}
\end{eqnarray}
where,

\begin{eqnarray}
&&h^A_{\mu}=g_3\cot\theta (\bar{t}\gamma_{\mu}\frac{\lambda^A}{2}t+\bar{b}_L\gamma_{\mu}\frac{\lambda^A}{2}b_L)+
\nonumber\\
&&+g_3\tan\theta(\bar{b}_R\gamma_{\mu}\frac{\lambda^A}{2}b_R+\sum_i\bar{q}_i\gamma_{\mu}\frac{\lambda^A}{2}q_i)
\label{curr443}
\end{eqnarray}

Here the sum goes over the fermions of the first two generations. This type of quark assignment with respect to the group $SU(3)_1\times SU(3)_2$ is not free from anomalies and requires
the addition of other fermion states to cure that. In the next section we will show how one can preserve the number of fermions of the standard model and get the correct hierarchy of the quark masses in a particular version of the top color model.

\section{An extended top condensate model}

We consider the top color interaction defined in the preceding section but applied only to the light quarks (up, down and  strange) and the top one.

We are interested in the four quark interactions which by color quark condensation can give masses however tiny to the light quarks. Specifically the term,
\begin{eqnarray}
&&-\frac{g_3^2\tan^2\theta}{M_B^2}( \sum_i\bar{q}_i\gamma^{\mu}\frac{\lambda^C}{2}q_i)(\sum_i\bar{q}_i\gamma^{\mu}\frac{\lambda^C}{2}q_i)
\nonumber\\
&&\approx\frac{g_3^2\tan^2\theta}{M_B^2}\frac{1}{2}\bar{q}_iq_i \bar{q}_iq_i
\label{res5554}
\end{eqnarray}
leads a mass of the light quark of the type:

\begin{eqnarray}
m_i=-\frac{g_3^2\tan^2\theta}{2M_B^2}\langle \bar{q}_iq_i \rangle
\label{maas5546}
\end{eqnarray}

\begin{figure}
\begin{center}
\epsfxsize = 4 cm
 \epsfbox{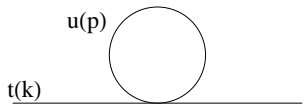}
\end{center}
\caption[]{%
A  possible diagram with top quark loop contributing to the quark masses.
}
\label{t4444}
\end{figure}

Note that as opposed to the Fermi interaction this mass term is positive. Thus the coloron interaction can give masses to the light quarks depending on the  corresponding diquark condensates. We shall consider the $SU(3)$ limit where the up, down and strange quark condensates have the same value \cite{Jora1}: $\langle \bar{q}_iq_i\rangle=-0.016$ $GeV^3$.
These masses are very small due to two factors: we expect $M_B\geq 246$ GeV and $\tan \theta\ll 1$. However in the presence of these masses however small the four fermion coloron interaction with the $\bar{q}$, $q$, $\bar{t}$ and t quarks and with top quark loop as in Fig. \ref{t4444} gives  significant contribution  to the light quark masses and should be regarded as an effective coupling of the light fermions with the Higgs doublet. We assume that a similar mechanism works in conjunction with the four quark components of the Higgs triplet this time coupled with  four light quarks through dimension eight effective operators with large coefficients.

It is hard to determine these couplings simply from the parameters of a strong dynamic theory so first we will try to establish some qualitative picture. For that we first list the quark massees as taken from \cite{PDG}:
\begin{eqnarray}
&&m_u=2.3^{+0.7}_{-0.5}\,\, MeV
\nonumber\\
&&m_d=4.8^{+0.5}_{-0.3}\,\, MeV
\nonumber\\
&&m_s=95\pm 5\,\, MeV
\nonumber\\
&&m_c=1.275\pm0.025 \,\,GeV
\nonumber\\
&&m_b=4.18\pm 0.03\,\, GeV
\nonumber\\
&&m_t=173.07\pm 0.52\pm 0.72\,\, GeV.
\label{res5534435}
\end{eqnarray}

The light quarks up, down and strange quarks can be considered in first order as it is usual in low energy QCD of the same mass.  This means that as in our initial set-up they are all triplets under $SU(3)_1$ and singlets with respect to  $SU(3)_2$. The masses of the charm and bottom quarks  are close but not small. We shall adopt here the point of view that these two quarks have a similar behavior with respect to the groups $SU(3)_1\times SU(3)_2$ with the left and right handed states divided between the two subgroups. The final assignment,
\begin{eqnarray}
&&(u_L,u_R,d_L,d_R,s_L,s_R,b_R,c_L) \,\,\,{\rm are }\,\,{\rm in}\,\, (3,1)\,\,{\rm of}\,\, SU(3)_1\times SU(3)_2
\nonumber\\
&&(t_L,t_R,b_L,c_R)\,\,\,{\rm are}\,\,{\rm in}\,\,(1,3)\,\,{\rm of}\,\,SU(3)_1\times SU(3)_2,
\label{asign5665}
\end{eqnarray}
is dictated by the hierarchy of the quark masses but it is not free from anomalies. The anomaly cancelation  would require the addition of extra fermion states that we shall not discuss here.

The neutral coloron current in Eq. (\ref{curr443}) will become:
\begin{eqnarray}
&&h_{\mu}^A=g_3\cot\theta(\bar{t}\gamma_{\mu}\frac{\lambda^A}{2}t+\bar{b}_L\gamma_{\mu}\frac{\lambda^A}{2}b_L+\bar{c}_R\gamma_{\mu}\frac{\lambda^A}{2}c_R)+
\nonumber\\
&&g_3\tan\theta(\bar{b}_R\gamma_{\mu}\frac{\lambda^A}{2}b_R+\bar{c}_L\gamma_{\mu}\frac{\lambda^A}{2}c_L+\sum_i\bar{q}_i\gamma_{\mu}\frac{\lambda^A}{2}q_i)
\label{res4435}
\end{eqnarray}
where the sum $\sum_i$ goes over the light quarks (up, down, strange).

In consequences the composite quark states in  Eqs. (\ref{tr77564}) and (\ref{full6575}) will be modified.
The Higgs doublet will be:
\begin{eqnarray}
&&\Phi_0=x_1t_L^{\dagger}t_R+x_2t_L^{\dagger}c_R
\nonumber\\
&&\Phi^+=x_2b_L^{\dagger}t_R +y_2 b_L^{\dagger}c_R
\label{res4435}
\end{eqnarray}

Here $x_i$ and $y_i$ are generic numbers to illustrate the fact that it is possible that the above structure corresponds to two Higgs doublets.

The Y=2 Higgs triplet is modified to:
\begin{eqnarray}
&&\chi^{++}=y_1b_L^{\dagger}t_R b_L^{\dagger}t_R+z_1b_L^{\dagger}c_R b_L^{\dagger}c_R+p_1b_L^{\dagger}t_R b_L^{\dagger}c_R
\nonumber\\
&&\chi^{+}=y_2b_L^{\dagger}t_Rt_L^{\dagger}t_R+z_2b_L^{\dagger}c_R t_L^{\dagger}t_R+p_2b_L^{\dagger}t_Rt_L^{\dagger}c_R+r_2b_L^{\dagger}c_Rt_L^{\dagger}c_R
\nonumber\\
&&\chi^0=y_3t_L^{\dagger}t_Rt_L^{\dagger}t_R+z_3t_L^{\dagger}c_Rt_L^{\dagger}c_R+p_3t_L^{\dagger}c_Rt_L^{\dagger}t_R
\label{h8878}
\end{eqnarray}
whereas the Y=0 Higgs triplet is:

\begin{eqnarray}
&&\xi^+=v_1b_L^{\dagger}t_R(t_L^{\dagger}t_R)^{\dagger}+w_1b_L^{\dagger}c_R(t_L^{\dagger}c_R)^{\dagger}+q_1b_L^{\dagger}c_R(t_L^{\dagger}t_R)^{\dagger}+
k_1b_L^{\dagger}t_R(t_L^{\dagger}c_R)^{\dagger}
\nonumber\\
&&\xi^0=v_2t_L^{\dagger}t_R(t_L^{\dagger}t_R)^{\dagger}+w_2t_L^{\dagger}c_R(t_L^{\dagger}c_R)^{\dagger}+q_2t_L^{\dagger}c_R(t_L^{\dagger}t_R)^{\dagger}+
k_2t_L^{\dagger}t_R(t_L^{\dagger}c_R)^{\dagger}
\nonumber\\
&&\xi^{-}=(\xi^+)^{\dagger}.
\label{res554646}
\end{eqnarray}

Here $y_i$, $z_i$, $p_i$, $r_i$, $w_i$, $q_i$, $k_i$ are generic coefficients.

For simplicity we wrote these states in terms of their $SU(2)_L$ and Y structures. Note that they may arrange themselves in multiple Higgs doublets and triplets  in accordance to their structure.

The final picture is that of an extended version of the standard model with multiple Higgs doublets and triplets. In the actual conjecture it is hard if not impossible to determine which scalar state corresponds to the Higgs like particle with a mass of 126 GeV found at the LHC. This topic is more complex and deserves a separate work. However by composition the model gives almost degenerate masses to the light quarks (up, down and  strange), higher masses the charm and bottom quarks and even higher mass to the top.

\section{Discussion and conclusions}

In essence we have the standard model of elementary particles with an extra gauge group $SU(3)_1\times SU(3)_2$ where $SU(3)_2$ is strong. This group is broken down to the color group $SU(3)_c$ by some mechanism and eight gauge bosons receive the mass  $M_B$.  For scales lower than $M_B$ these heavy bosons can be integrated out which leads to four fermion interactions. Some of the quarks in the model like the top quark, left handed bottom and right handed charm have the correct quantum numbers with respect to the strong group $SU(3)_2$ to permit the appearance of composite states and of strong vacuum condensates. These states form multiple Higgs doublets, which are two quark states and triplets with a four quark structure.

We then assume that five of the quarks (or all of them) form chiral QCD condensate with the same value such that one has a chiral $SU(5)_L\times SU(5)_R$ broken down to the diagonal part.
Through the four fermion interaction these quarks receive tiny masses as in:
\begin{eqnarray}
&&m_i\approx-\frac{g_3^2\tan^2\theta}{M_B^2}\langle \bar{q}q\rangle
\nonumber\\
&&m_b\approx-\frac{g_3^2}{M_B^2}\langle \bar{q}q\rangle
\nonumber\\
&&m_c\approx-\frac{g_3^2}{M_B^2}\langle \bar{q}q\rangle.
\label{m5546}
\end{eqnarray}
where $m_i$ refers to the three light quarks.

Then through diagrams as that depicted in Fig.\ref{t4444} the quarks couple with the scalar condensates in the model. This would correspond to effective couplings of two fermion states with the
Higgs doublet and of four fermion states with the components of the Higgs triplets.

We can estimate only the relative contribution of these condensates to the masses:
\begin{eqnarray}
&&m_i'\approx g_2^3 X
\nonumber\\
&&m_b'\approx g_3^2 \frac{\cot(\theta)}{\sqrt{2}}X
\nonumber\\
&&m_c'\approx g_3^2 \frac{\cot(\theta)}{\sqrt{2}}X
\nonumber\\
&&x m_t\approx g_3^2 \cot^2(\theta) X
\label{contr66657}
\end{eqnarray}

Here X correspond to an effective vacuum average expectation value in the model and  the proportionality factors are estimated from the square of the operators. The mass of the top quark
has contribution from both two and four quark vacuum condensates and $x=\frac{-1+\sqrt{5}}{2}$  measures the two top quark condensate contribution (see \cite{Jora} for details).

Note that the quark assignment considered in the present version of the model may lead to flavor changing neutral currents among the lightest quarks.  However even by taking into account the current experimental limits on these which set  a lower bound on the coloron mass one can still achieve the correct mass for the top quark  provided that either the top coupling constant of the two top quark composite state is very small or the mass of that state is very high (see \cite{Jora}).  Both cases eliminate the two top quark composite state as a candidate for the Higgs boson with a mass of $125-126$ GeV found at the LHC.

\begin{figure}
\begin{center}
\epsfxsize = 10cm
 \epsfbox{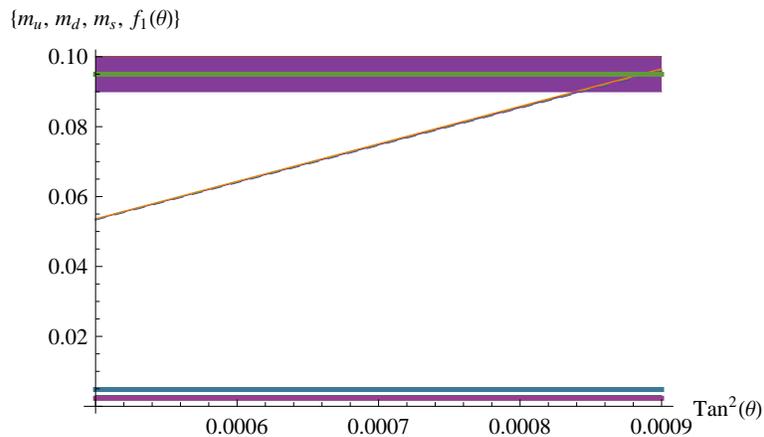}
\end{center}
\caption[]{%
Plot of the function $f_1(\theta)$ (yellow line) as function of $\tan^2(\theta)$. The light quark masses, $m_u$ (purple), $m_d$ (blue) and $m_s$ (green and purple) are represented as horizontal bands.
}
\label{tt1}
\end{figure}

\begin{figure}
\begin{center}
\epsfxsize = 10 cm
 \epsfbox{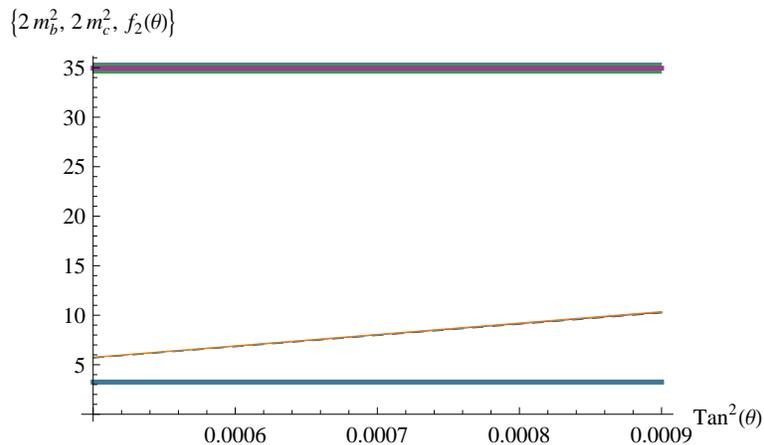}
\end{center}
\caption[]{%
Plot of the function $f_2(\theta)$ (yellow line) as function of $\tan^2(\theta)$. The heavy quark function of the masses $2m_c^2$ (blue), $2m_b^2$ (purple) are represented as horizontal bands.

}
\label{tt2}
\end{figure}

 We can rewrite the content of Eq. (\ref{contr66657}) as,
 \begin{eqnarray}
 &&m_i=\tan^2(\theta)(x m_t)=f_1(\theta)
 \nonumber\\
 &&2m_b^2=2m_c^2=\tan^2(\theta) (x m_t)^2=f_2(\theta),
 \label{rez456}
 \end{eqnarray}
in order to get rid of the unwanted X factor.

In Fig. \ref{tt1} we plot the function $f_1(\theta)$ as compared to the light fermion masses to show that for values of $\tan^2\theta\in (0.0005,0.0008)$ there are solutions that correspond to an average light quark mass in the interval  $0.06-0.095$ GeV. In Fig. \ref{tt2} we plot the function $f_2(\theta)$ to obtain that for the same  interval for $\tan^2(\theta)$ there are solutions that interpolate between the two heavy quarks masses. Thus there is an acceptable solution for $\theta$ that agrees with Eq. (\ref{rez456}). Moreover this solution also satisfies the requirement $\tan\theta\ll 1$ of a topcolor theory.


The model is very complicated and has a complex dynamics which may justify further corrections to the masses.  A next step would be to consider a possible mixing between the composite Higgs multiplets existent in the model.

The problem of leptons masses and that of the Higgs particles in this framework will receive an extensive treatment elsewhere.

\section*{Acknowledgments} \vskip -.5cm
The work of R. J. was supported by PN 09370102/2009 and by a grant of the Ministry of National Education, CNCS-UEFISCDI, project number PN-II-ID-PCE-2012-4-0078. The work of J. S. was supported in part by the US DOE under Contract No. DE-FG-02-85ER 40231.


\begin{thebibliography}{15}



\bibitem{Lindner} W. A. Bardeen, C. T. Hill and M. Lindner, Phys. Rev. D {\bf 41}, 1647 (1990).
\bibitem{Hill} C. Hill, Phys. Lett. B {\bf 266}, 419-424 (1991).
\bibitem{Simmons} C. T. Hill, E. H. Simmons, Phys. Rept. {\bf 381}, 235 (2003).
\bibitem{Hill3} B. A. Dobrescu and C. T. Hill, Phys. Rev. Lett. {\bf 81}, 2634-2637 (1998).
\bibitem{Miransky} V. A. Miransky, M. Tanabashi and K. Yamawaki, Phys. Lett.B {\bf 221}, 177 (1989); Mod. Phys. Lett. {\bf 4}, 1043 (1989).
\bibitem{Nambu} Y. Nambu, EFI-89-09 (1989).
\bibitem{Marciano} W. J. Marciano, Phys. Lett. {\bf 62}, 2793 (1989).
\bibitem{Hill2} C. Hill, Phys. Lett. B {\bf 345}, 483-489 (1995); arXiv:hep-ph/9411426.
\bibitem{Hill3} R. S. Chivukula, B. A. Dobrescu, H. Georgi, C. T. Hill, Phys. Rev. D {\bf 59}, 075003 (1999); arXiv:hep-ph/9809470.
\bibitem{Jora} R. Jora, S. Nasri and J. Schechter, Phys. Rev. D {\bf 87}, 115027 (2013); arXiv:1304.7139.
\bibitem{Jora1} A. H. Fariborz, R. Jora and J. Schechter, Phys. Rev. D {\bf 77}, 094004 (2008); arXiv:0801.2552.
\bibitem{Georgi} H. Georgi and M. Machacek, Nucl. Phys. B {\bf 262}, 463-477 (1985).
\bibitem{PDG} J. Beringer \textit{et al} (Particle Data GRoup), Phys. Rev. D {\bf 86}, 010001 (2012).
\end{thebibliography}
\end{document}